\documentclass{article}
\usepackage[T1]{fontenc}
\usepackage{microtype,graphicx,booktabs,array,tabularx}
\usepackage{amsmath,amssymb,mathtools,textcomp,xurl}
\usepackage{hyperref}
\usepackage[preprint]{icml2026}
\usepackage{placeins}

\newcolumntype{Y}{>{\raggedright\arraybackslash}X}
\newcommand{\Normal}{\mathcal N}
\newcommand{\E}{\mathbb E}
\newcommand{\Var}{\operatorname{Var}}
\newcommand{\KL}{\operatorname{KL}}

\graphicspath{{figures/common174/}{figures/focused-draft/}}
\makeatletter
\renewcommand{\ICML@preprint}{\textit{Preprint.}}
\makeatother
\hypersetup{pdftitle={Inferring Neural Activity Changes under Recording Uncertainty with Transformers and Variational Inference},pdfauthor={Nilson Chapagain and Pooja Chapagain},pdfsubject={arXiv preprint},hidelinks}
\icmltitlerunning{Neural Activity Changes under Recording Uncertainty}
\begin{document}
\twocolumn[
\icmltitle{Inferring Neural Activity Changes under Recording Uncertainty with Transformers and Variational Inference}
\begin{icmlauthorlist}
\icmlauthor{Nilson Chapagain}{tamu}
\icmlauthor{Pooja Chapagain}{ulm}
\end{icmlauthorlist}
\icmlaffiliation{tamu}{Department of Statistics, Texas A\&M University, College Station, Texas, USA}
\icmlaffiliation{ulm}{Department of Biology, University of Louisiana at Monroe, Monroe, Louisiana, USA}
\icmlcorrespondingauthor{Nilson Chapagain}{nchapagain@tamu.edu}
\icmlkeywords{Longitudinal electrophysiology, neuron tracking, recording uncertainty, in-context learning, variational inference, count models}
\vskip 0.3in
]
\printAffiliationsAndNotice{}
\begin{abstract}
A difference in recorded firing rate across days may reflect a change in neuronal activity, missed spikes or recording a different neuron. We present a statistical approach that combines neuron tracking with Bayesian estimates of stationary mean recorded firing rates. Bounds under assumed differences in spike detection connect recorded rates to biological rate changes. We assess recording loss and uncertainty in rate changes in simulations, and count-model fit in recordings. We also compare two transformer estimators with Gaussian and flow variational inference under a shared temporal negative-binomial model. Flow variational inference agrees most closely with reference posteriors, while predictive performance varies across mice. Rate-change intervals contain true simulated changes less often than intended. In UnitMatch recordings from five mice, some inferred changes remain supported when spike detection is allowed to differ between days; others depend on equal detection. These interpretations require a correct neuron match. The approach identifies which inferred rate changes remain supported under specified recording assumptions and where uncertainty limits interpretation.
\end{abstract}
\section{Introduction}\label{sec:intro}
Repeated recordings from the same brain region are used to study changes in neuronal firing during learning and over time~\citep{vanbeest2025,huang2026,rule2019}. We call each period of data collection a \emph{recording session}. During a session, a probe measures electrical signals, and spike sorting groups detected spikes into \emph{units}, each intended to represent one neuron. Comparing firing rates across days requires following the same neuron across sessions. It also requires allowing for changes in \emph{spike capture}, the fraction of that neuron's spikes detected and assigned to its unit. A difference in recorded rate can reflect any of these sources: a change in firing, an incorrect match or a change in spike capture.

An incorrect match can change what we infer about neuronal responses. Within a recording, errors in spike assignment can bias tuning estimates or create apparent correlations between neurons~\citep{wood2008,ventura2009,pillow2013}. Across recordings, tools such as UnitMatch compare the average shapes of a unit's spikes to estimate the probability of a match~\citep{vanbeest2025}. DANT groups units across sessions and corrects for probe motion~\citep{huang2026}. Where the true identities are unknown, agreement between tracking procedures provides evidence for a match~\citep{barnett2016}. We compare UnitMatch with concatenated sorting, which sorts both recording days together. Agreement supports a proposed match, although errors in the two procedures may be related because they use the same recordings. \citet{lu2026} used the same five-mouse release to benchmark tracking and studied firing-rate changes in separate macaque data. We build on tracking by asking which firing-rate comparisons retain support when we vary assumptions about missed spikes and count variation.

Even when the neuron and fraction of spikes captured are fixed, the observed count varies over time. \citet{stevenson2011} simulated fixed movement tuning to show how spiking variability can produce apparent changes in tuning. We ask a related question about mean firing rates. Changes in neuronal excitability can make spike counts more variable than a Poisson model allows and can correlate counts in neighboring time intervals~\citep{goris2014}. Latent dynamical models account for unobserved population activity~\citep{pandarinath2018}, and Stiefel Manifold Dynamical Systems allow the mapping from that activity to measurements to vary across trials~\citep{lee2026}. For comparisons of mean firing rates, we therefore examine both the count model and the method used to calculate uncertainty.

We present a statistical approach for inferring changes in mean neuronal firing rate from recordings made on different days. We analyze spike times and average spike waveforms, the shapes of the recorded electrical signals, from the five-mouse UnitMatch dataset~\citep{unitmatchdata2024,vanbeest2025}. For each mouse, established tracking methods link groups of recorded spikes across two recording days. A proposed match contains one group from each day that may come from the same neuron. We call this a \emph{candidate pair}. For each group, we estimate the stationary mean recorded firing rate: the expected number of spikes per second, averaged over the model's rate fluctuations. The mean over the particular observed window is a separate quantity. Because spikes may be missed, we specify how much the fraction detected and assigned to each group could differ between days. The analysis then reports which increases or decreases in the neuron's mean rate remain supported over that range, together with the matching evidence and uncertainty in the recorded rates. This interpretation assumes that both groups represent the same neuron and that spikes from other neurons contribute negligibly (Section~\ref{sec:sensitivity}).

To quantify uncertainty in each recording's mean firing rate, we combine its observed spike counts with a prior distribution, which describes parameter uncertainty before observing the recording. The resulting posterior distribution describes uncertainty after observing the counts. A credible interval contains a specified probability under this posterior. Variational inference (VI) fits an approximation to that posterior by optimization~\citep{blei2017}. Structured approximations can account for relationships between unobserved states at different times~\citep{archer2015}. Prior-data fitted networks (PFNs) offer another route: they learn from simulated datasets and use a new set of observations to estimate parameter distributions~\citep{muller2022,hollmann2025,reuter2025,mittal2025}. Neural posterior estimation (NPE) learns parameter distributions from simulated examples. NPE-PFN uses a pretrained PFN to estimate conditional distributions for individual parameters and combine them into a joint posterior~\citep{vetter2025}. MetaBeta estimates posteriors for hierarchical models, where groups share information through common parameters, and can refine them with additional sampling~\citep{kipnis2026}.

We compare NPE-PFN~\citep{vetter2025}, PFN-NPE~\citep{pickens2026}, Gaussian VI and flow VI under the same negative-binomial (NB) time-series model. PFN-NPE estimates the posterior from numerical features extracted by a PFN. Gaussian VI uses a Gaussian approximation, while flow VI transforms Gaussian draws through invertible functions to represent a more flexible distribution. NB counts allow greater variance than their mean, and the temporal model allows nearby counts to be related. In simulations with known rates, we assess interval coverage, the fraction of intervals containing the generating value~\citep{modrak2023}. We also assess how much the intervals narrow relative to the prior and how well the methods predict the next count. For selected recordings, we compare their posterior draws with numerical reference samples that pass the sampling checks in Section~\ref{sec:forecast-methods}. These comparisons assess whether each approximation gives accurate and informative uncertainty estimates for rate comparisons.

For the mouse recordings, the temporal NB model estimates changes in mean recorded rate. A separate Gaussian autoregressive (AR) model relates each count to the preceding count. We use it to check sensitivity to the prior and to the time window included in the analysis. Together, these analyses provide separate checks on the recording assumptions, the count model and the posterior approximation used to infer a change.

Tables~\ref{tab:related-work-comparison} and~\ref{tab:inference-work-comparison} separate the recording analyses from the posterior estimators. We reuse established methods to connect tracking evidence, posterior rate contrasts and bounds under missed-spike assumptions. Our contribution is the evaluated rate-change analysis, including separate checks of count-model fit and posterior approximation.

\begin{table*}[!t]
\centering
\caption{Comparison with neural-recording analyses. Entries describe the cited formulations and evaluations.}
\label{tab:related-work-comparison}
\begingroup
\fontsize{8}{9.6}\selectfont
\setlength{\tabcolsep}{2pt}
\renewcommand{\arraystretch}{1.1}
\begin{tabularx}{\linewidth}{@{}>{\raggedright\arraybackslash}p{0.18\linewidth}>{\raggedright\arraybackslash}p{0.19\linewidth}*{5}{>{\centering\arraybackslash}X}@{}}
\toprule
\textbf{Work} & \textbf{Recordings} & \textbf{Cross-day tracking} & \textbf{Activity target} & \textbf{Uncertain spike labels} & \textbf{Missed-spike rate bounds} & \textbf{Temporal model}\\
\midrule
\citet{vanbeest2025}\newline UnitMatch & Mouse Neuropixels & \(\checkmark\) & Responses & --- & --- & ---\\
\addlinespace[2pt]
\citet{huang2026}\newline DANT & Mouse/rat Neuropixels & \(\checkmark\) & Responses & --- & --- & ---\\
\addlinespace[2pt]
\citet{lu2026} & Macaque arrays; mouse benchmark & \(\checkmark\) & Rates; tuning & --- & --- & ---\\
\addlinespace[2pt]
\citet{dickey2009} & Macaque Utah arrays & \(\checkmark\) & Rates; tuning & --- & --- & ---\\
\addlinespace[2pt]
\citet{stevenson2011} & Macaque reaching recordings & --- & Tuning & --- & --- & Dynamic tuning\\
\addlinespace[2pt]
\citet{wood2008} & Macaque waveforms and movement & --- & Tuning & Propagated & --- & ---\\
\addlinespace[2pt]
\citet{carlson2014} & Rat waveforms across days & \(\checkmark\) & Counts; rates & Joint model & --- & Period-specific rates\\
\addlinespace[2pt]
\citet{pandarinath2018}\newline LFADS & Macaque/human cortical recordings & --- & Single-trial rates & --- & --- & Nonlinear dynamics\\
\addlinespace[2pt]
\citet{lee2026}\newline SMDS & Primate spikes; rodent calcium imaging & --- & Population activity & --- & --- & Linear dynamics; changing emissions\\
\midrule
\textbf{This work} & \textbf{Five-mouse Neuropixels} & Reused trackers & \textbf{Stationary mean rates} & --- & \(\boldsymbol{\checkmark}\) & \textbf{NB and AR models}\\
\bottomrule
\end{tabularx}
\par\smallskip
\raggedright
Checks indicate reported analyses; a dash denotes another target or an analysis not reported in the cited formulation. Spike-label uncertainty concerns propagation of assignment uncertainty within a recording, rather than neuron identity across days. Our biological-rate bounds condition on a correct match and an assumed range for relative spike capture. Period-specific rates do not imply serial dynamics. LFADS aligns populations without tracking individual neurons. Histograms of times between spikes and autocorrelation features used for matching are not temporal activity models.
\par\endgroup
\end{table*}

\begin{table*}[tp]
\centering
\caption{Comparison with transformer posterior estimators. The reported checks use different models and datasets, rather than a common performance benchmark.}
\label{tab:inference-work-comparison}
\begingroup
\fontsize{8}{9.6}\selectfont
\setlength{\tabcolsep}{2pt}
\renewcommand{\arraystretch}{1.1}
\begin{tabularx}{\linewidth}{@{}>{\raggedright\arraybackslash}p{0.17\linewidth}>{\raggedright\arraybackslash}p{0.21\linewidth}*{5}{>{\centering\arraybackslash}X}@{}}
\toprule
\textbf{Work} & \textbf{Estimator and models} & \textbf{Frozen TabPFN} & \textbf{VI comparison} & \textbf{Posterior reference} & \textbf{Posterior checks} & \textbf{Predictive checks}\\
\midrule
\citet{reuter2025} & Transformer with flow; regression, factors and mixtures & --- & Gaussian; flow & Analytic; sampling & Posterior distances & Responses\\
\addlinespace[2pt]
\citet{vetter2025}\newline NPE-PFN & Conditional posteriors; general and neural simulators & \(\checkmark\) & Published VI results & Benchmark samples & Calibration\newline concen\-tration & Neural simulations\\
\addlinespace[2pt]
\citet{pickens2026}\newline PFN-NPE & PFN features with flow; static and dynamic simulators & \(\checkmark\) & --- & Samples; grids & Dispersion; region size & ---\\
\addlinespace[2pt]
\citet{kipnis2026}\newline MetaBeta & Prior-conditioned posteriors; hierarchical regression & --- & ADVI; Pathfinder & Sampling & Coverage; posterior scale & Leave-one-out\\
\midrule
\textbf{This work} & \textbf{Four estimators; temporal NB counts} & \(\checkmark\) & \textbf{Gaussian; flow} & \textbf{Sampling} & \textbf{Coverage; interval width} & \textbf{Next-bin counts}\\
\bottomrule
\end{tabularx}
\par\smallskip\raggedright
Symbols follow Table~\ref{tab:related-work-comparison}; reported checks do not by themselves establish calibration or superiority. NPE-PFN's VI entry uses published pyloric-simulator results. ADVI is automatic differentiation variational inference. Our PFN routes reuse pretrained TabPFN; the posterior estimators themselves are established methods.\par
\endgroup
\end{table*}

\section{Data and firing-rate comparisons}\label{sec:data}
\subsection{Five-mouse dataset and analysis cohort}
The UnitMatch example dataset contains two recording days for each of five mice~\citep{unitmatchdata2024,vanbeest2025}. The released spike times come from sorting each session separately and sorting the two sessions together. The release also includes average spike waveforms, the electrical signal shapes around detected spikes, for each half of each recording. Probe geometry specifies the positions of the recording channels. Unit-quality summaries describe each unit over a stated time window. We retain 1,584 unit-sessions with a good cluster-table quality label from recordings whose spike timestamps span the analysis window and the following ten-second forecast window. Individual units need not fire near both ends of the window. Each unit-session represents one sorted unit in one recording. Table~\ref{tab:five-recordings} lists the eligible units and recording identifiers. We use the released spike-event arrays and align separate-session and concatenated outputs by recording identifiers and timestamp offsets.

For each recording, we divide the 29-minute window $[60,1800)$ seconds after its start into 174 ten-second bins. We count recorded spikes in each bin. We use elapsed time from the timestamps as the nominal exposure. The timestamps do not establish how long the probe was actively recording or what the animal was doing throughout that time. The unit-quality summaries refer to time windows that may be shorter than 29 minutes. When waveform features are unavailable, the unit can still contribute spike counts.

\begin{table*}[tp]\centering\small
\caption{Recordings and eligible sorted units. Dates are followed by the released sorting-folder identifier in parentheses. All recordings use Probe0. Each unit-session is one sorted unit in one recording.}\label{tab:five-recordings}
\setlength{\tabcolsep}{4pt}
\begin{tabular}{@{}lllrlrr@{}}\toprule
Mouse & Animal & First recording & Units & Second recording & Units & Total\\\midrule
Mouse1 & AL032 & 2019-11-21 (1) & 278 & 2019-11-22 (1) & 270 & 548\\
Mouse2 & JF067 & 2022-02-14 (4) & 100 & 2022-02-15 (5) & 75 & 175\\
Mouse3 & AV008 & 2022-03-12 (IMRO\_1) & 64 & 2022-03-13 (IMRO\_1) & 48 & 112\\
Mouse4 & EB019 & 2022-07-21 (1) & 40 & 2022-07-22 (1) & 46 & 86\\
Mouse5 & CB016 & 2021-09-28 (1) & 355 & 2021-09-29 (1) & 308 & 663\\\midrule
Total & & & 837 & & 747 & 1,584\\\bottomrule
\end{tabular}
\end{table*}

\subsection{Recorded and biological firing-rate ratios}
Consider a candidate pair with one sorted unit from each day. For day $d\in\{0,1\}$, let $\lambda_d>0$ be the neuron's stationary mean firing rate and $m_d$ its stationary mean recorded rate, both per nominal second. These means average over the fluctuations described by that day's model. Assuming the units represent the same neuron and contamination from other neurons is negligible, define $g_d=m_d/\lambda_d$, with $0<g_d\le1$, as the fraction of generated spikes recorded and assigned to the unit under the same averaging. Then
\begin{equation}\label{eq:recorded}
 \begin{aligned}
 m_d&=g_d\lambda_d,\\
 R&=\frac{m_1}{m_0}
   =\underbrace{\frac{g_1}{g_0}}_{q}
    \underbrace{\frac{\lambda_1}{\lambda_0}}_{B}.
 \end{aligned}
\end{equation}
The recorded-rate ratio $R$ is the product of relative capture $q$ and the biological stationary mean-rate ratio $B$. Capture may depend on activity: periods with more generated spikes contribute more to $g_d$. We do not adjust for differences in behavior between days. The mean over the latent rates that occurred during a particular window is a separate target, defined in Eq.~\eqref{eq:window-rate}.

For example, $R=0.35$ is compatible with $(B,q)=(0.35,1)$, a biological rate decrease under equal capture, or $(B,q)=(1,0.35)$, unchanged biological firing with reduced capture. Both interpretations assume the units represent the same neuron. The identity map $M$ specifies which units are proposed to match across days. We check these assignments before interpreting the rate ratios.

\subsection{Spike capture and change criteria}\label{sec:sensitivity}
From $B=R/q>0$, assumed capture bounds $q\in[L,U]$, with $0<L\le U$, give
\begin{equation}\label{eq:capture-envelope}
 \frac{R}{U}\le B\le\frac{R}{L}.
\end{equation}
For a central 90\% credible interval $[a,b]$ for $R$, with $a>0$, the corresponding bounds for $B$ are $[a/U,b/L]$. We consider equal capture, $L=U=1$, and illustrative ranges $[L,U]=[0.8,1.25]$ and $[0.5,2]$. These bounds are assumed, not estimated from the recordings.

A pair meets the directional change criterion if $a/U>1.2$ for an increase or $b/L<1/1.2$ for a decrease. The 1.2-fold margin is descriptive, with no simultaneous error guarantee across pairs or validated biological threshold. Interval availability means that the interval passes the model and numerical reporting checks; it does not by itself establish directional change. For the Gaussian AR model, reporting requires checks on the probability of nonpositive means and on numerical precision (Section~\ref{sec:ar-settings}).

\begin{figure*}[tp]\centering
\includegraphics[width=0.80\linewidth]{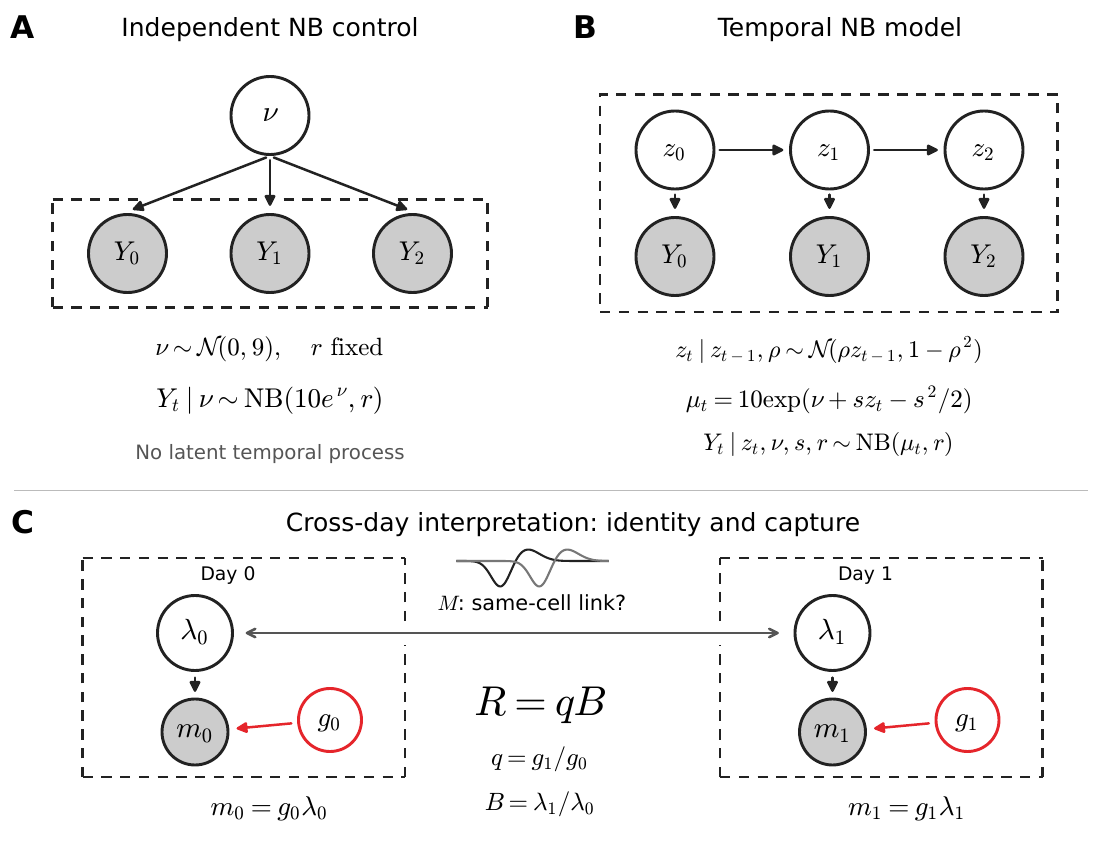}
\caption{Spike-count models and firing-rate comparisons. (A) Independent negative-binomial (NB) counts. (B) NB state-space model with latent state $z_t$. Both models use ten-second bins and set capture to one. (C) For the proposed match $M$, the recorded-rate ratio is $R=qB$, where $q$ is relative spike capture and $B$ is the biological firing-rate ratio.}\label{fig:problem}
\end{figure*}

Figure~\ref{fig:problem} summarizes our count models and rate comparisons~\citep{lee2026}.

\section{Statistical methods and recording checks}\label{sec:methods}
The analysis follows each candidate pair from tracking to a rate comparison. We first check whether the units could come from the same neuron and how missed spikes could change the comparison. We then model the count variation within each recording to estimate the uncertainty in its mean rate.

\subsection{Cross-session unit tracking and within-session controls}\label{sec:matching}
Before matching units across days, we test how well their spike waveforms identify them within one recording. We extract features from each recording half using the released code and probe geometry (82 time samples and 384 channels per half). Units with the same sorting label in both halves serve as presumed matches. We compare waveform shape alone with shape and spatial position. In five-fold cross-validation, we estimate the matching thresholds on the training folds. For each held-out unit, we test two sets of ten possible matches from the other recording half. One set contains the presumed match and nine alternatives. The other contains ten alternatives, allowing us to test whether the procedure rejects all of them. The waveform score is centered-waveform correlation. The combined score subtracts $0.5(d_{\rm centroid}/100)^2$, where $d_{\rm centroid}$ is the distance between amplitude-weighted waveform centers in released channel-coordinate units. For each training unit, we randomly draw ten units with different labels and record their largest score. The cutoff is the empirical 95th percentile of these maxima across training units, rounded up to an observed value. We accept the highest-scoring held-out comparison only if its score exceeds the cutoff.

For matching across days, UnitMatchPy 3.6.1 combines six waveform-similarity scores and corrects for probe movement, using two metric/drift iterations and default score smoothing~\citep{vanbeest2025}. The initial analysis includes units with the waveform features needed for UnitMatch and our within-session comparison. UnitMatchPy gives two probabilities for each pair, one from the first day to the second and one in the reverse direction. We accept a pair when both probabilities exceed 0.5 and its mean probability is uniquely highest among the alternatives for each unit. We repeat matching after restricting this set to units with the released waveform-quality label GoodID=1. This checks sensitivity to the set of eligible units. This restriction changes the competing matches, the score distributions and the prior match probabilities. The count analysis uses the separate quality labels in the cluster tables.

We check the waveform matches against \emph{concatenated sorting}, which sorts both days' recordings together. In each session, we link separately sorted units to jointly sorted units by counting their shared spike events. Ambiguous events remain in the denominators of the overlap fractions. We vary the spike-pairing tolerance over 0, 1 and 3 samples (0, 33.3 and 100 microseconds at 30 kHz). We also vary the minimum overlap fraction over 0.5, 0.7 and 0.9. The overlap fraction is the proportion of one unit's spike events also assigned to the other unit. A link must meet the threshold in both directions and be the unique largest overlap for both units. A separately sorted unit from each day linked to the same jointly sorted unit forms a candidate pair. We keep candidates from any of the nine settings or either waveform analysis. We call a pair \emph{tracking-concordant} when the initial UnitMatch analysis and concatenated sorting both support it.

\subsection{Spike-thinning controls}
Even for the same unit, missing more spikes on one day can create an apparent rate change. We randomly split a unit's recorded spikes into subsets A and B. We then retain each B spike with probability $q$. Here $q$ is the imposed relative capture. This control removes spikes independently from recorded, sorted spike trains and leaves cell assignments fixed. The variables $A$, $B_{\rm kept}$ and $B_{\rm dropped}$ count spikes assigned to A, kept in B and removed from B. Conditional on the original count $N$,
\begin{equation}\label{eq:loss}
\begin{gathered}
 (A,B_{\rm kept},B_{\rm dropped})\mid N\\
 \sim\operatorname{Multinomial}\{N;0.5,0.5q,0.5(1-q)\}.
\end{gathered}
\end{equation}
We compare three assumed capture ranges: $L=U=1$ (equal capture), $L=U=q$ (known capture), and $[L,U]=[0.5,2]$ (bounded capture). Each subset uses a Poisson count likelihood, 870 seconds of nominal exposure (half the original), and a $\operatorname{Gamma}(0.5,0.1)$ recorded-rate prior, with shape and rate arguments. Gaussian VI approximates the log recorded-rate posterior; we then adjust $R$ using the assumed capture bounds. An exact Gamma--beta calculation checks the posterior probability behind each directional criterion. For an equal-tailed 90\% interval, this probability must exceed 0.95 beyond the corresponding change margin. For each $q\in\{1,0.8,0.5\}$, we generate 100 resamples conditional on each of the 1,584 observed spike trains.

\subsection{Count models and blocked cross-validation}
Firing-rate uncertainty also depends on how much spike counts vary within a recording. Poisson regression assumes equal conditional mean and variance. Negative-binomial (NB) regression allows the variance to exceed the mean, a property called overdispersion. With mean $\mu>0$ and size $r>0$, its probability mass function is
\begin{equation}\label{eq:nb}
\begin{aligned}
 p(y\mid\mu,r)&=\frac{\Gamma(y+r)}{\Gamma(r)\Gamma(y+1)}\\
 &\quad\times\left(\frac{r}{r+\mu}\right)^r
 \left(\frac{\mu}{r+\mu}\right)^y,
 \quad y=0,1,\ldots,
\end{aligned}
\end{equation}
with $\E(Y)=\mu$ and $\Var(Y)=\mu+\mu^2/r$. At a fixed mean, a smaller $r$ gives greater count variance.

For each count distribution, the time-varying model has intercept $\beta_0$ and cosine coefficients $\beta_1,\ldots,\beta_6$:
\begin{equation}\label{eq:count-regression}
\begin{aligned}
 \log\mu_t&=\log10+\beta_0\\
 &\quad+\sum_{k=1}^{6}\beta_k\cos\{k\pi(t+0.5)/T\}.
\end{aligned}
\end{equation}
Here $t=0,\ldots,T-1$ and $T=174$. The offset $\log10$ accounts for ten-second bins. The constant-rate model omits the sum. We use independent priors $\beta_0\sim\Normal(0,9)$ and $\beta_k\sim\Normal(0,1)$, $k=1,\ldots,6$, with variance as the second argument, and fit a full-covariance Gaussian variational approximation. To test prediction within a recording, we divide its counts into six consecutive 290-second blocks, numbered 0--5. The three outer folds hold out block pairs (0,3), (1,4) and (2,5). We exclude the two bins (20 seconds) adjacent to each held-out block from training. Two inner folds alternate the training blocks and retain the two-bin exclusion. Their pooled predictive log loss, the negative log probability assigned to the held-out counts, selects $r\in\{0.5,1,3,10,30\}$. We refit at the selected $r$ before scoring the outer test blocks. The outer-fold scores evaluate the full procedure, including selection of $r$. Because training can use observations on either side of a held-out block, these scores measure interpolation within a recording. The NB comparison was designed after inspecting the Poisson results and is exploratory. Section~\ref{sec:numerical} describes the optimization and quadrature.

\subsection{Negative-binomial state-space model}\label{sec:temporal-model}
The NB count regressions allow overdispersion, but their residuals remain correlated across nearby bins. We represent this dependence with a latent state $z_t$ that follows a first-order autoregressive process and determines the expected count. For a recording with $T$ bins, the states and recorded counts $Y_t$ satisfy
\begin{equation}\label{eq:temporal}
 \begin{aligned}
 z_0&\sim\Normal(0,1),\\
 z_t\mid z_{t-1},\rho&\sim\Normal(\rho z_{t-1},1-\rho^2),\\
 \mu_t&=e_t d_t\exp(\nu+s z_t-s^2/2),\\
 Y_t\mid z_t,\nu,s,r&\sim\operatorname{NB}(\mu_t,r).
 \end{aligned}
\end{equation}
The transition applies for $t=1,\ldots,T-1$. Temporal persistence is $\rho\in(-1,1)$, with $\operatorname{Corr}(z_t,z_{t-1}\mid\rho)=\rho$. The scale $s>0$ sets the size of log-rate fluctuations, and $r>0$ controls NB count variance. We set nominal exposure and detection to $(e_t,d_t)=(10,1)$ to estimate the rate of recorded spikes. Real detection loss enters separately through the capture bounds. At fixed global parameters, $z_t\mid\rho\sim\Normal(0,1)$ for every $t$. Hence $\E\{\exp(s z_t-s^2/2)\}=1$ and $\E(Y_t)=10e^\nu$. Thus $e^\nu$ is the stationary mean recorded rate per nominal second, corresponding to $m_d$ in Eq.~\eqref{eq:recorded}.

The global parameters $\nu,s,\rho,r$ stay fixed within a recording, while the latent state $z_t$ varies between bins. We assign them independent priors,
\begin{equation}\label{eq:priors}
 \begin{gathered}
 \nu\sim\Normal(0,9),\quad s\sim\operatorname{HalfNormal}(0.5),\\
 (\rho+1)/2\sim\operatorname{Beta}(2,2),\\
 \log r\sim\Normal(\log3,1).
 \end{gathered}
\end{equation}
The second argument of each Normal distribution is its variance. The Half-Normal prior has scale 0.5. The parameter $\nu$ is the log of the stationary mean recorded rate, which averages over the state distribution. The log mean rate over a particular recording instead averages over the states that occurred in its $T$ bins,
\begin{equation}\label{eq:window-rate}
 \log\left\{T^{-1}\sum_{t=0}^{T-1}
 \exp(\nu+s z_t-s^2/2)\right\}.
\end{equation}
Joint draws of the states $z_{0:T-1}$ and parameters give uncertainty intervals for this finite-window rate. Averaging the log rates is different: their stationary expectation is $\nu-s^2/2$.

\subsection{Variational posterior approximations}\label{sec:vi-inference}
The rate comparisons use the joint posterior of the four global parameters. VI estimates this distribution together with the latent states. Let $D$ denote the observed counts and $\theta=(\nu,s,\rho,r,z_{0:T-1})$ collect the parameters and latent states. Here $z_{0:T-1}$ denotes the sequence of states from bin 0 to bin $T-1$. Their joint posterior is
\begin{equation}\label{eq:posterior}
 p(\theta\mid D)\propto p(D\mid\theta)p(\theta).
\end{equation}
We fit the variational density $q_\eta$, parameterized by $\eta$, by maximizing the evidence lower bound (ELBO)~\citep{blei2017},
\begin{equation}\label{eq:elbo}
 \begin{aligned}
 \mathcal L(\eta)&=\E_{q_\eta}\{\log p(D,\theta)-\log q_\eta(\theta)\},\\
 \KL(q_\eta\Vert p(\cdot\mid D))&=\log p(D)-\mathcal L(\eta).
 \end{aligned}
\end{equation}
The Kullback--Leibler divergence $\KL(q_\eta\Vert p(\cdot\mid D))$ measures the discrepancy from the posterior. In Eq.~\eqref{eq:elbo}, $\log p(D)$ is constant in $\eta$. Maximizing the ELBO therefore minimizes that discrepancy without evaluating the marginal likelihood.

We compare two variational families. A full-covariance Gaussian represents correlations among parameters and states. A normalizing flow transforms Gaussian draws through invertible functions to represent a more flexible density. Our flow uses four inverse-autoregressive transforms, each with two 512-unit hidden layers~\citep{kingma2016}. Both methods parameterize the latent states through innovations, the new random fluctuations at each time step, and rescale these during optimization. We run 12,000 updates from each of three starts. We select the fit with the highest independently estimated ELBO. Section~\ref{sec:numerical} gives the fitting budget and the transformation that preserves the target posterior. We assess posterior accuracy separately through simulation and reference comparisons.

\subsection{Transformer-based posterior estimation}
The transformer methods approximate the joint posterior of $\nu,s,\rho,r$ after the latent states have been integrated out. Their simulated examples pair generating parameter values with recordings from the same model and prior. The context is a set of these examples supplied to the pretrained transformer when it processes a new recording. The two methods differ in how they turn these examples into a posterior: one estimates conditional densities in sequence, and the other fits a conditional flow.

NPE-PFN uses the pretrained TabPFNv2 regressor with fixed weights (TabPFN library version 2.0.9)~\citep{vetter2025}. To draw from the joint approximation, it samples the transformed coordinates $u=(\nu,\log s,\log\{(1+\rho)/(1-\rho)\},\log r)$ in order. Each draw is conditional on the observed counts $D$, the earlier coordinates in the same draw and the simulated examples in context $\mathcal C$,
\begin{equation}\label{eq:pfn}
 q_\phi(u\mid D,\mathcal C)
 =\prod_{j=1}^{4}q_\phi(u_j\mid u_{<j},D,\mathcal C).
\end{equation}
Here $u_{<j}$ denotes the coordinates sampled before coordinate $j$, and $\phi$ denotes the posterior estimator's parameters. The density $q_\phi$ is the resulting joint approximation. For each recording, we select the 1,000 simulated examples whose 174 log-transformed counts are nearest to the observed counts. We standardize each time-bin coordinate using its mean and standard deviation across the 11,000 context and training simulations, then measure Euclidean distance. Validation examples are excluded. We take 4,096 draws without importance reweighting and evaluate densities in batches of 512. Section~\ref{sec:numerical} describes the checks on sampling and density evaluation.

PFN-NPE uses the fixed TabPFNv2 regressor to represent each recording by numerical features (TabPFN library version 6.3.2)~\citep{pickens2026}. Each global parameter has a separate context containing the first 1,000 simulated examples. Each parameter-specific estimator produces 192 features for the recording. We concatenate the 768 features and reduce them to 64 principal components fitted on the 10,000 training examples. A conditional neural spline flow, a flexible density estimator, estimates the posterior from these features~\citep{durkan2019}. It uses five invertible spline transforms, two hidden layers of 128 units and eight intervals in each spline. The same principal components and posterior estimator serve all 174-bin recordings.

Both methods use a pool of 13,000 simulated recordings and transform counts as $y\mapsto\log(1+y)$ while keeping them in time order. NPE-PFN selects examples at inference time, while PFN-NPE fits the feature reduction and flow in advance. These steps give the methods different fitting and inference costs. Their global-parameter draws suffice for stationary rate comparisons. Forecasting also needs the state at the end of the observed window, which we estimate separately.

\begin{figure*}[tp]\centering
\includegraphics[width=0.82\linewidth]{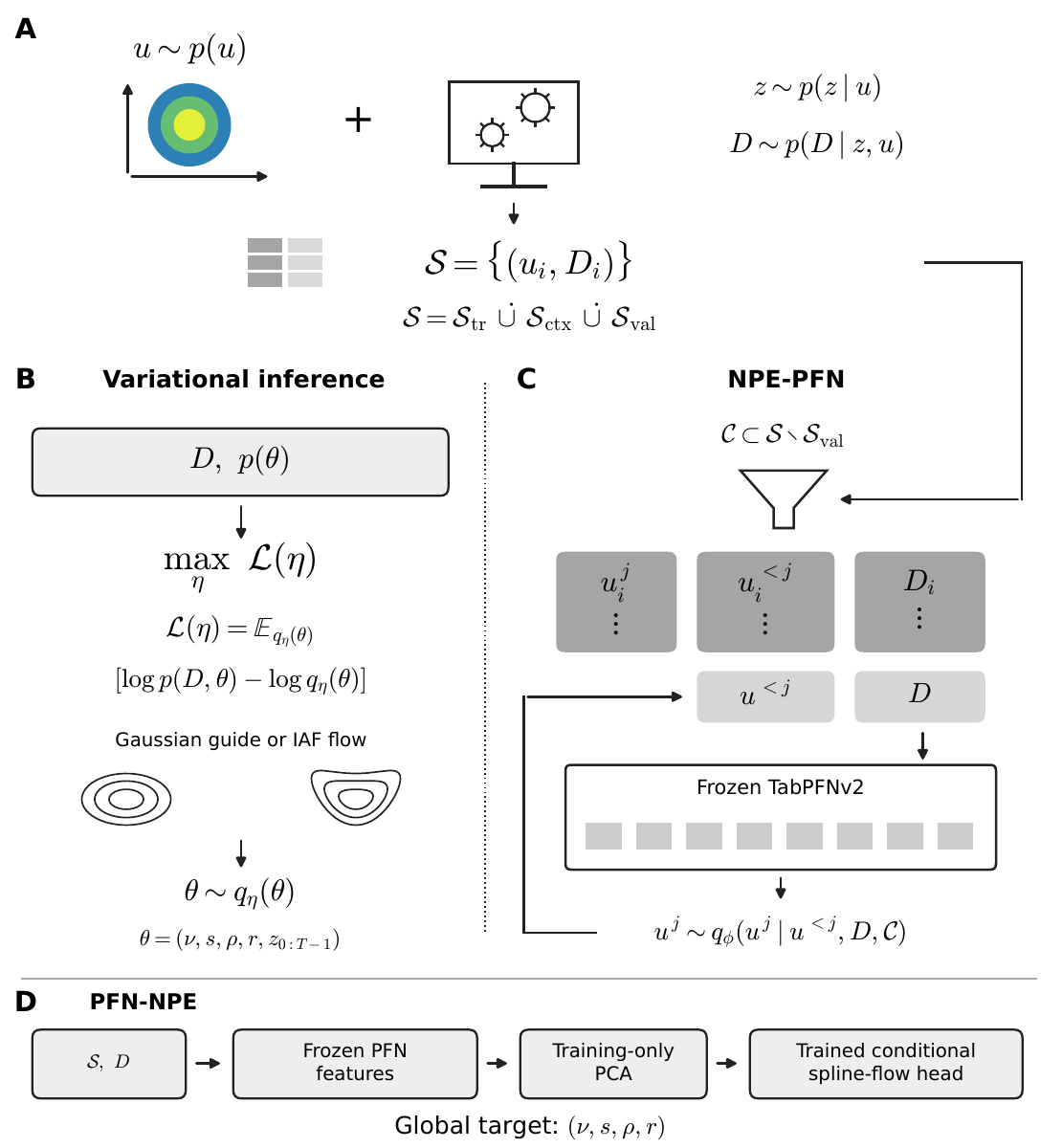}
\caption{Simulation and posterior estimation. (A) $\mathcal S$ pairs transformed parameters $u_i$ with simulated count series $D_i$. Subscripts $\mathrm{tr}$, $\mathrm{ctx}$ and $\mathrm{val}$ indicate training, fixed context and validation subsets. (B) VI approximates the joint posterior of parameters and states. (C) NPE-PFN samples parameters sequentially for observed counts $D$ using context $\mathcal C$. (D) PFN-NPE fits a conditional flow to reduced frozen TabPFNv2 features.}\label{fig:interface}
\end{figure*}

Figure~\ref{fig:interface} summarizes our posterior-estimation routes~\citep{vetter2025}.

\subsection{Gaussian autoregressive model assessment}\label{sec:ar-settings}
The temporal NB comparison holds the likelihood and prior fixed across posterior estimators. For the mouse data, we assess sensitivity to the prior and the analysis window. The shorter window contains only time bins covered by both units' quality summaries. We call it the quality-overlap window. We use a Gaussian AR working model for those checks. Its stationary mean count per bin is $m$, and $v>0$ is the innovation variance,
\begin{equation}\label{eq:gaussian-ar-current}
\begin{aligned}
Y_0&\sim\Normal\{m,v/(1-\rho^2)\},\\
Y_t\mid Y_{t-1}&\sim\Normal\{m+\rho(Y_{t-1}-m),v\}.
\end{aligned}
\end{equation}
The Gaussian model estimates a mean recorded rate of $m/10$, the counterpart of $e^\nu$ under the temporal NB model. The two models use different likelihoods and priors. The Gaussian model allows negative counts and nonpositive means, which matter when taking a log ratio. For each session, a log-ratio interval requires $\Pr(m\le0\mid D)\le10^{-6}$ and $\min_k m^{(k)}>0$, where $m^{(k)}$ are the sampled mean counts. The interval endpoints must also pass the numerical-refinement check. We apply these rules under two priors, first to all 174 bins and then to the bins covered by both units' quality summaries. The overlap must have at least 30 bins. When the windows coincide, there is only one distinct window. We integrate $m,v$ analytically conditional on $\rho$, retaining the stationary initial density, then integrate over $\rho$. The primary priors are $\rho\sim\operatorname{Uniform}(-0.95,0.99)$ and $v\sim\operatorname{InverseGamma}(2,1)$. The alternative uses $\rho\sim\operatorname{Uniform}(-0.99,0.999)$ and $v\sim\operatorname{InverseGamma}(1,0.1)$. Both use $m\mid v\sim\Normal(0,v/10^{-4})$; inverse-Gamma arguments are shape and scale. We intersect quality-summary windows with $[60,1800)$ seconds and round inward to ten-second bins. Quadrature and interval checks are specified in Section~\ref{sec:numerical}.

\subsection{Posterior contrasts of firing rates}\label{sec:common174-change}
For a candidate match, let the left and right recordings refer to the first and second days. We fit them separately. Under Eq.~\eqref{eq:temporal}, the stationary mean recorded rate is $e^\nu$, so the log ratio from the first day to the second is
\begin{equation}\label{eq:delta-nu}
 \Delta\nu=\nu_{\rm right}-\nu_{\rm left}=\log R.
\end{equation}
From 4,096 draws per session, we obtain
\begin{equation}\label{eq:contrast-interval}
\begin{gathered}
 [a_\Delta,b_\Delta]=\\
 Q_{(0.05,0.95)}
 \left(\{\nu_{\rm right}^{(j)}-\nu_{\rm left}^{(i)}\}_{i,j=1}^{4096}\right).
\end{gathered}
\end{equation}
Here $Q$ returns the indicated linearly interpolated empirical quantiles of all $4,096^2$ equally weighted differences. This uses the product of the two empirical marginal posteriors. For observed recordings, it assumes posterior independence between sessions conditional on the proposed match. In simulations we use the fixed recording pairs $(0,1),(2,3),\ldots,(126,127)$, giving 64 independent pairs with known contrasts.

For the finite-window rate in Eq.~\eqref{eq:window-rate}, we also need the latent states from each recording. Gaussian and flow VI provide joint draws of the states and parameters for this calculation. We therefore assess the finite-window contrast for VI and the stationary contrast for all four methods. To account for differential spike capture, we use the bounds $L\le q\le U$ from Section~\ref{sec:sensitivity}. A stationary log-ratio interval $[a_\Delta,b_\Delta]$ then gives the bounds $[a_\Delta-\log U,\,b_\Delta-\log L]$ on the log ratio of biological stationary mean rates under the same-cell and capture assumptions.

\subsection{Reference inference and forecasting}\label{sec:forecast-methods}
For numerical reference samples, we use the No-U-Turn Sampler (NUTS)~\citep{hoffman2014} to sample the global parameters and latent states jointly. Before inspecting outcomes, we select one unit-session per recording by the smallest SHA-256 hash of its identifier. Each of four chains uses 6,000 warmup and 4,000 retained draws, target acceptance 0.999, maximum tree depth 12 and a full mass matrix. We accept a reference fit only when it passes fixed sampling checks. With four chains (separately initialized sampling runs), we require the rank-normalized statistic $\widehat R\le1.01$ and $\min(\mathrm{ESS}_{\rm bulk},\mathrm{ESS}_{\rm tail})\ge400$, where ESS is effective sample size. Each chain must have an estimated Bayesian fraction of missing information of at least 0.3. No divergences are allowed, and fewer than 1\% of retained draws may reach the tree-depth limit. ESS accounts for dependence between samples~\citep{vehtari2021}; divergences, the energy diagnostic~\citep{betancourt2016} and the tree-depth limit check different sampling difficulties. Fits passing these checks provide numerical references, not exact posterior distributions. We compare 4,096 approximate draws with all 16,000 retained reference draws. We report how many fits qualify and retain the others in the denominator.

Agreement with a reference posterior and prediction answer different questions. For prediction, we use the observed counts $D_{0:173}=(Y_0,\ldots,Y_{173})$ to forecast $Y_{174}$, the next ten-second count in $[1800,1810)$ seconds. VI and NUTS give joint draws of the parameters and the last observed-window state $z_{173}$. The transformer methods give parameter draws. For each global-parameter draw, we process the observed counts in time order to obtain the conditional distribution of $z_{173}$. This filtering step estimates the latent state at the end of the observed window. We then propagate this state distribution through one further transition and the NB observation model to predict the next count. The joint factorization is
\begin{equation}\label{eq:forecast-factor}
 \begin{gathered}
 q(\nu,s,\rho,r\mid D_{0:173})\,\\[-1pt]
 {}\times p(z_{173}\mid\nu,s,\rho,r,D_{0:173})\,\\[-1pt]
 {}\times p(Y_{174}\mid z_{173},\nu,s,\rho,r).
 \end{gathered}
\end{equation}
Averaging this distribution over the parameters and $z_{173}$ gives the forecast. The parameter draws already condition on the observed counts, so we give them equal weight. We check the numerical integration before forecasting and use $Y_{174}$ only to score the forecast. See Section~\ref{sec:numerical} for the precision criteria and Monte Carlo replication.

\subsection{Numerical integration and fitting}\label{sec:numerical}

For the Poisson and NB count regressions, we fit a Gaussian approximation with L-BFGS-B~\citep{byrd1995} and Newton-type refinement. L-BFGS-B uses at most 2,000 iterations and 50 line-search steps, with gradient tolerance $10^{-9}$ and function tolerance $10^{-14}$. Let $C$ be the lower-triangular covariance Cholesky factor and $f$ the fitting objective. We initialize $C=0.1I$, retrying with $C=0.01I$ if needed, where $I$ is the identity matrix. Fits require finite $f,\nabla f$, $\|\nabla f\|_\infty\le10^{-6}$ and $\min_j C_{jj}>10^{-10}+10^{-12}$. The optimizer's lower bound is $C_{jj}\ge10^{-10}$. Gauss--Hermite refinement must change the NB objective and gradient by at most $10^{-7}$ and predictive log masses and CDFs by at most $10^{-5}$. Difficult Poisson predictive integrals use more nodes or adaptive integration.

For temporal VI, standard innovations are $\epsilon_0=z_0$ and $\epsilon_t=(z_t-\rho z_{t-1})/\sqrt{1-\rho^2}$. Optimization uses $w_0=s^{0.4}\epsilon_0$ and $w_t=s^{0.4}(1-\rho^2)^{0.15}\epsilon_t$. At fixed $s,\rho$, the log Jacobian from $w$ to $z$ is $0.35(T-1)\log(1-\rho^2)-0.4T\log s$. We retain it and the Normal normalization constants. Adam uses 32 draws per update, learning rate 0.003, multiplicative decay 0.9999 and gradient clipping at 100~\citep{kingma2015}. Checkpoints at 4,000, 8,000 and 12,000 updates are compared using 2,048 independent ELBO draws; the selected fit is assessed with 8,192 fresh draws. Objectives, ELBO contributions and retained draws must be finite, with parameters in their stated support. These checks do not certify posterior accuracy.

The transformer bank contains 1,000 context, 10,000 training and 2,000 validation recordings, separate from evaluation data. NPE-PFN uses eight TabPFN estimators, temperature 0.9 and probability averaging after softmax. For 32 draws, sampled and recomputed log densities must agree to absolute tolerance 0.001 and relative tolerance $10^{-5}$, using the same 512-row query batch. PFN-NPE trains with batches of 256, gradient clipping at 5 and initial learning rate 0.0005, following a cosine schedule for at most 200 epochs. Validation loss selects the checkpoint, with early-stopping patience 20. The PCA and parameter standardization use training examples only. Each recording yields 4,096 joint parameter draws. The retained transformed $\nu$ coordinates permit rate contrasts even when converting extreme draws back to the original scales causes rounding or overflow.

For Gaussian AR inference, Gauss--Legendre quadrature is refined from 256 to 512, 1,024 or 2,048 nodes. Successive grids must agree within $10^{-7}$ for the mean-count posterior CDF at its refined-grid 5th, 50th and 95th percentiles, the log normalizer and the posterior mean of $\rho$. For posterior moments, let $\Delta$ denote the refined-grid minus preceding-grid value. We require
\begin{equation}\label{eq:ar-moment-precision}
\begin{aligned}
 \frac{|\Delta\E(m\mid D)|}{1+|\E_{\rm fine}(m\mid D)|}&\le10^{-7},\\
 \frac{|\Delta\Var(m\mid D)|}{1+\Var_{\rm fine}(m\mid D)}&\le10^{-7}.
\end{aligned}
\end{equation} Two scrambled four-dimensional Sobol sequences use $2^{15}$ points and a $2^{14}$-point refinement~\citep{scipysobol}. No draw is discarded or truncated to ensure positivity. Central 90\% log-ratio endpoints must agree within 0.005 log units across scrambles and refinements. Difference endpoints use tolerance 0.005 times interval width plus $10^{-6}$ count units.

Forecasting uses 4,096 paths per Monte Carlo replicate. VI uses joint parameter and last-state draws; NUTS uses 1,024 evenly spaced draws per chain. PFNs select 512 global draws without replacement and eight conditional last-state draws per global draw. All paths propagate to $z_{174}$. With $p_{bj}$ denoting the conditional NB probability of the next count on path $j$ of replicate $b$, we estimate
\begin{equation}\label{eq:forecast-probability}
 \widehat p_b(Y_{174}\mid D_{0:173})
 =\frac1{4096}\sum_{j=1}^{4096}p_{bj}.
\end{equation}
Global draws have equal weight. We apply the negative logarithm within each replicate and average the eight losses. When only log loss passes its checks, we evaluate these masses without generating future counts.

The PFN conditional filter uses 1,024 Gauss--Legendre nodes on $[-10,10]$. We test 64 fixed parameter draws against 2,048 nodes on $[-10,10]$ and $[-12,12]$. Integration and omitted predictive mass use tolerance $10^{-8}$, and normalization uses $10^{-10}$. Last-state CDF and quantile tolerances are 0.002 and 0.05. Predictive CDF, quantile and relative-mean tolerances are 0.001, one count and 0.001. NB probability recurrence must agree within 0.005 log units, and zero-count CDF arithmetic within 0.001. Numerical retries preserve the parameter weights, seeds, observations and likelihood.

\section{Experimental design and evaluation}\label{sec:design}
\subsection{Simulations for the Gaussian autoregressive model}
To test the Gaussian AR model with dependent spike counts, we generate counts with NB marginals using a Gaussian copula~\citep{cario1996},
\begin{equation}\label{eq:copula}
\begin{aligned}
 z_t&=\rho z_{t-1}+\sqrt{1-\rho^2}\epsilon_t,\\
 Y_t&=F_{\rm NB}^{-1}\{\Phi(z_t)\}.
\end{aligned}
\end{equation}
Here $z_0,\epsilon_t$ are standard Normal variables, $\Phi$ is the standard Normal cumulative distribution function (CDF) and $F_{\rm NB}$ is the NB CDF with $r=3$. The inverse CDF converts each latent value into a count while preserving the NB marginal distribution. The simulations vary latent persistence $\rho\in\{0,0.5,0.9\}$ and use mean counts of $(20,20)$ or $(20,30)$ across the two recordings. We use the same latent-noise draws for both mean settings.

We compare coverage and availability of Gaussian AR intervals for differences and log ratios of stationary mean counts. The checks use 256 pairs per combination of observation model, $\rho$ and mean counts, with 174 bins per recording. Gaussian observations have stationary variance approximately 153.33; NB-copula observations use $r=3$. A further 256 pairs are generated from the primary Gaussian prior. We report both interval availability and coverage conditional on availability.

\subsection{Temporal-model calibration design}
We draw parameters from the priors in Eq.~\eqref{eq:priors} and simulate 128 recordings from the temporal NB model, independently of the transformer simulation bank. For each global parameter, we estimate prior-averaged coverage of central 50\%, 80\%, 90\% and 95\% credible intervals and divide their widths by the corresponding prior widths. We assess the joint posterior with the energy score, which compares posterior draws with the generating parameter vector; lower scores are better~\citep{gneiting2007}. For this score we standardize the coordinates $u$ from Eq.~\eqref{eq:pfn} using
\begin{equation}\label{eq:standardized-u}
 \widetilde u_j=
 \frac{u_j-\E_{\rm prior}(u_j)}
      {\sqrt{\Var_{\rm prior}(u_j)}},
 \qquad j=1,\ldots,4,
\end{equation}
where prior moments are taken under Eq.~\eqref{eq:priors}. Coverage is also averaged over this prior-generated parameter distribution~\citep{modrak2023}, with Wilson binomial confidence intervals~\citep{wilson1927}. Equal-tail intervals use linearly interpolated quantiles and closed endpoints. We compute the energy score using the mean of the 4,096 distances to the generating vector minus half the mean distance in 2,048 disjoint posterior-draw pairs. All methods use the same pairing indices. Paired 95\% percentile bootstrap intervals use 10,000 resamples of recordings.

We form 64 independent pairs in a fixed order to assess coverage of $\Delta\nu$. We also assess coverage of finite-window log-rate differences for methods that provide joint state and parameter draws. We report coverage and interval widths for all simulation pairs.

\subsection{Posterior and predictive accuracy}
We use separate metrics for parameter uncertainty and prediction. For selected reference recordings, we compare samples in the standardized coordinates $\widetilde u$. The sliced Wasserstein-2 distance is $\mathrm{SW}_2=\{256^{-1}\sum_{k=1}^{256}W_{2,k}^2\}^{1/2}$, where $W_{2,k}$ is the one-dimensional Wasserstein-2 distance between samples projected onto fixed unit direction $k$. We retain all draws in transformed coordinates, avoiding numerical overflow on the original parameter scales. For each projection, we integrate squared empirical-quantile differences over the pooled breakpoints. Passed sampling checks do not remove finite-sample error in these numerical references.

For each unit-session or simulated recording, let $\widehat p_b$ be the predictive probability estimate from Monte Carlo replicate $b$. We score the next-bin count by averaging eight log losses,
\begin{equation}\label{eq:forecast-loss}
 \widehat\ell=-\frac18\sum_{b=1}^{8}
 \log\widehat p_b(Y_{174}\mid D_{0:173}),
\end{equation}
Lower log loss indicates better prediction. We use log loss because extreme posterior draws can destabilize predictive means. The primary comparison uses unit-sessions scored by all four methods. We also compare each PFN with flow VI on recordings scored by both methods, summarize scores by session, and repeat the comparison after excluding the largest absolute differences in paired log loss. We treat unit-sessions within a mouse as dependent.

\section{Results}\label{sec:five-results}

\subsection{Posterior calibration and parameter recovery}
Flow VI had the lowest mean joint energy score across 128 simulations (0.843), followed by Gaussian VI (0.931), PFN-NPE (0.990) and NPE-PFN (1.086). The difference in mean energy score between flow VI and PFN-NPE was $-0.147$. Resampling the 128 recordings gave a 95\% percentile bootstrap interval of $[-0.207,-0.092]$ for this difference. All four methods yielded posterior draws for every simulation and all 1,584 unit-sessions.

\begin{figure*}[tp]\centering
\includegraphics[width=0.84\linewidth]{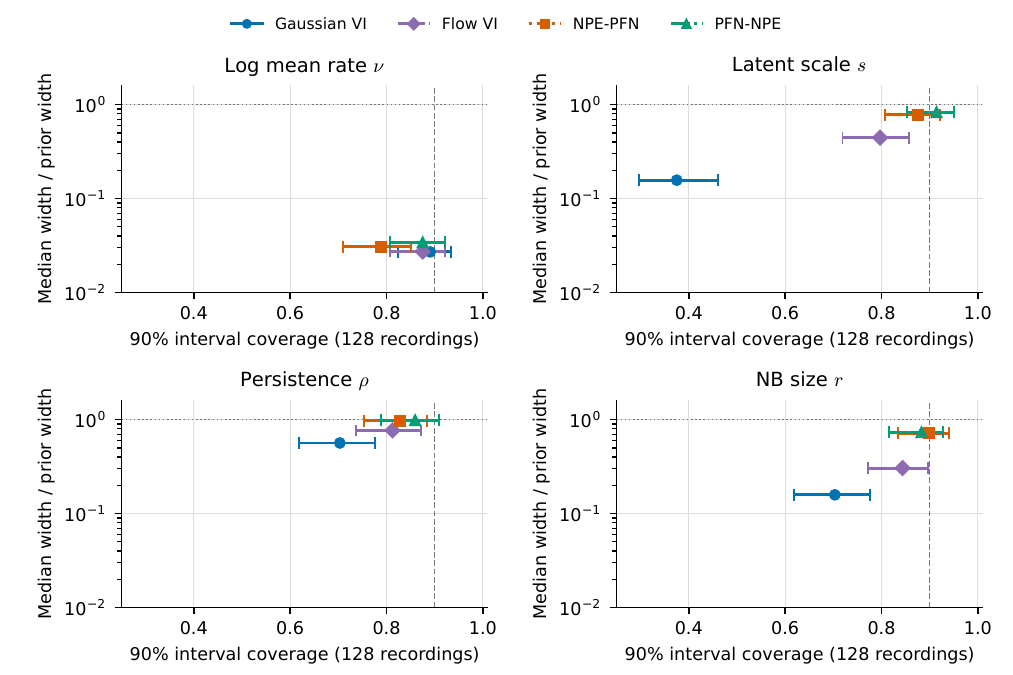}
\caption{Marginal 90\% credible-interval coverage and interval width relative to the corresponding prior interval in 128 simulations per method.}\label{fig:five-calibration}
\end{figure*}

\begin{figure*}[tp]\centering
\includegraphics[width=0.82\linewidth]{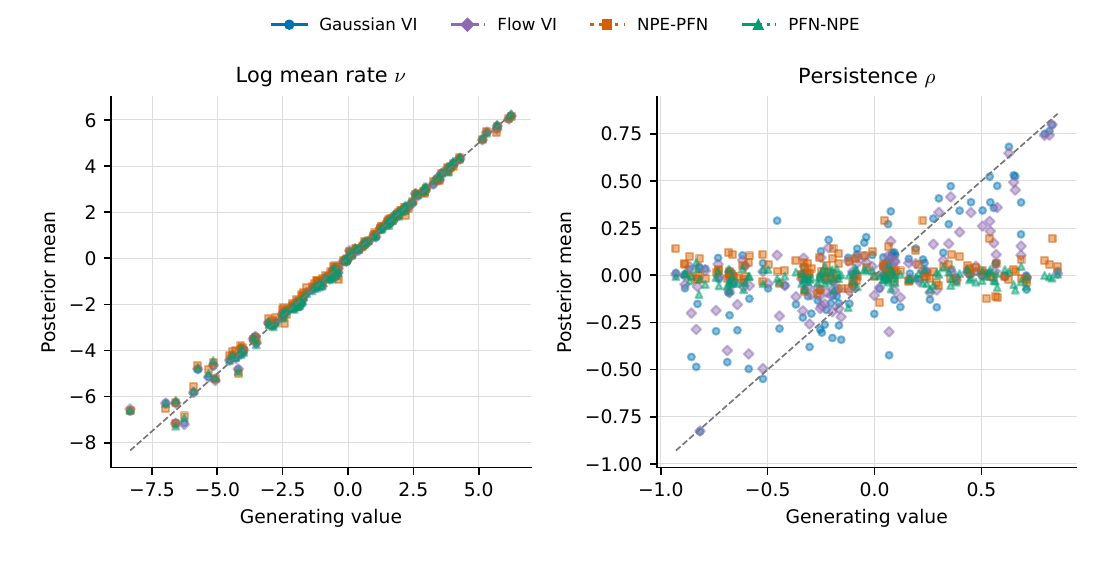}
\caption{Posterior means versus generating values for log stationary mean recorded rate $\nu$ and persistence $\rho$, the correlation between consecutive latent states, in 128 simulations per method.}\label{fig:parameter-recovery}
\end{figure*}

All four methods recovered $\nu$ (log stationary mean recorded rate) more closely than $\rho$ (temporal persistence). Flow VI improved marginal interval coverage for $s$ and $\rho$ relative to Gaussian VI. PFN-NPE had higher empirical coverage than flow VI for $s$ and $\rho$. Its estimates of $\rho$ stayed near the prior mean, with median intervals about 99\% as wide as the prior interval (Figs.~\ref{fig:five-calibration} and~\ref{fig:parameter-recovery}).

We could calculate rate contrasts from the retained draws of $\nu$ in every NPE-PFN fit. In seven fits, converting persistence draws back to $\rho$ rounded some values to $-1$ or $1$. Extreme draws also destabilized some posterior means on the original parameter scales (Section~\ref{sec:numerical}).

\subsection{Agreement with reference posteriors}
Seven of the ten NUTS reference fits to mouse recordings passed the sampling diagnostics. Flow VI had the smallest $\mathrm{SW}_2$ in each case. Mean $\mathrm{SW}_2$ values were 0.127 for flow VI, 0.188 for Gaussian VI, 1.321 for NPE-PFN and 1.332 for PFN-NPE.

\subsection{Coverage of posterior firing-rate contrasts}
\begin{figure*}[tp]\centering
\includegraphics[width=0.90\linewidth]{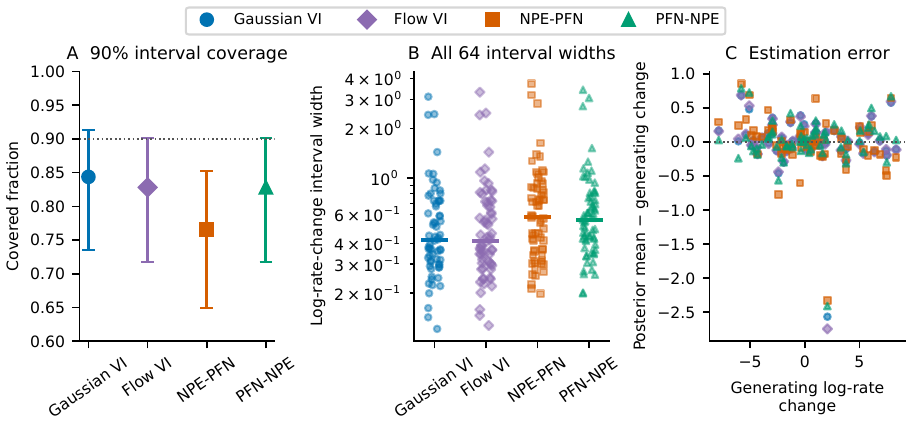}
\caption{Stationary log-rate contrasts $\Delta\nu=\nu_{\rm right}-\nu_{\rm left}$ in 64 simulated pairs. (A) Empirical coverage with 95\% Wilson confidence intervals. (B) Credible interval widths and medians. (C) Posterior mean minus generating contrast, plotted against the generating contrast.}\label{fig:five-changes}
\end{figure*}

To check uncertainty in rate changes, we paired independent simulated recordings with known stationary mean rates and estimated their log-rate differences $\Delta\nu$. Across 64 pairs, empirical coverage of nominal 90\% credible intervals for $\Delta\nu$ ranged from 76.6\% to 84.4\%. The 95\% Wilson confidence intervals for these estimated coverage rates included the nominal 90\% level for three of the four methods. The transformer methods had larger median interval widths than the VI methods (Fig.~\ref{fig:five-changes}). For the finite-window log-rate contrast, each VI method's 90\% credible intervals contained the generating contrast in 56/64 pairs. Unlike the stationary-rate comparison, the finite-window calculation requires joint state and parameter draws.

\subsection{One-step-ahead spike-count prediction}\label{sec:five-prediction}
Flow VI had the lowest mean log loss on the 123 simulations scored by all four methods (3.276; range across methods 3.276--3.305).

\begin{table*}[tp]\centering\small
\setlength{\tabcolsep}{3pt}
\caption{Mean one-step-ahead predictive log loss by mouse. Each row uses the same unit-sessions for all four methods. Lower values indicate better prediction.}\label{tab:five-forecast}
\begin{tabular}{@{}lrrrrr@{}}\toprule
Mouse & Unit-sessions scored & Gaussian VI & Flow VI & NPE-PFN & PFN-NPE\\\midrule
Mouse1 & 548 & 3.774 & 3.769 & 3.815 & 3.786\\
Mouse2 & 164 & 3.730 & 3.722 & 3.875 & 3.800\\
Mouse3 & 104 & 3.697 & 3.707 & 4.032 & 4.065\\
Mouse4 & 83 & 5.461 & 5.417 & 5.112 & 5.285\\
Mouse5 & 653 & 3.432 & 3.430 & 3.520 & 3.506\\
\bottomrule\end{tabular}

\end{table*}

Both PFNs had lower mean log loss than VI in Mouse4, while VI had the lowest mean in each of the other four mice (Table~\ref{tab:five-forecast}).

To check dependence on forecast availability, we also compared the 70 Mouse4 unit-sessions scored by all four methods before recomputing forecasts that failed the numerical checks. In this subset, both PFNs had lower mean log loss than flow VI in the 21 July recording and higher mean loss in the 22 July recording. Across these 70 unit-sessions, the median within-unit-session difference in log loss favored flow VI over both PFNs. The Mouse4 mean-log-loss advantage also held after excluding the four unit-sessions with the largest absolute differences from flow VI.

Across all methods and observed or simulated recordings, predictive log loss was available for 6,818 of 6,855 forecasts, including forecasts from NUTS reference fits. For 250 forecasts, the numerical checks supported reporting log loss but not predictive means or intervals. All 37 unavailable scores were from NPE-PFN (32 unit-sessions and five simulated recordings; Section~\ref{sec:numerical}).

\subsection{Recording controls and count-model fit}
Waveform shape and position recovered the presumed match in most evaluable within-recording tests. When the presumed match was absent, the procedure sometimes accepted a false match. Mouse4 had too few eligible units for the within-session validation.

The thinning control compares two subsets of spikes from the same unit, then removes spikes from one subset to impose a known recording loss. With no imposed loss, 109 of 158,400 resamples had 90\% intervals indicating an apparent rate change beyond the 1.2-fold margin under equal capture. One met this change criterion under $q\in[0.5,2]$. When each spike in subset B was removed with probability 20\%, 67,624 of 158,400 resamples met the change criterion when the analysis assumed equal capture. Accounting for the known loss reduced this count to 109. With 50\% thinning, the corresponding counts were 156,094 and 171. Under relative capture in $[0.5,2]$, 3 and 88 resamples met the criterion at 20\% and 50\% thinning, respectively.

Recorded counts also vary within a session, which affects the precision of its mean rate. The constant-rate NB model predicted held-out counts better than constant-rate Poisson regression in 1,562/1,584 unit-sessions. Allowing the log rate to vary over time through six cosine terms improved prediction over the constant-rate NB model in 123/1,584 unit-sessions. Model residuals remained correlated across nearby bins.

\begin{figure*}[tp]\centering
\includegraphics[width=0.84\linewidth]{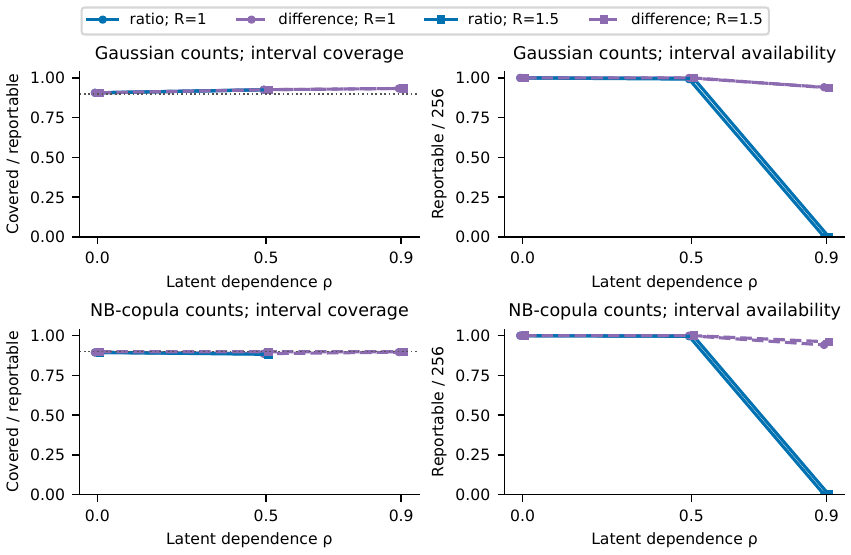}
\caption{Gaussian AR interval coverage (left) and availability (right) for 256 simulated pairs per setting under the primary prior. Coverage is conditional on an available interval. Circles and squares denote true ratios of 1 and 1.5, simulated with shared noise draws.}\label{fig:five-ar}
\end{figure*}

The simulations show which Gaussian AR intervals remain reportable as serial dependence increases. Most log-ratio intervals met the reporting criteria under moderate dependence. At the strongest setting ($\rho=0.9$), under the primary prior, rate-difference intervals met the criteria for 240--246 of 256 simulated pairs in each combination of observation model and true mean counts. No log-ratio intervals met the criteria (Fig.~\ref{fig:five-ar}).

\section{Cross-session firing-rate changes in the UnitMatch dataset}\label{sec:five-application}
We applied the rate comparisons to 371 candidate pairs identified by UnitMatch or concatenated sorting. Of these, 134 pairs had support from both tracking procedures.

For each proposed match, we compared the stationary mean recorded rates from the two days, using $R$ in Eq.~\eqref{eq:recorded}. All four methods produced posterior contrasts for all 371 pairs. Three NPE-PFN fits required using the retained draws of $\nu$ directly from the transformed parameterization (Section~\ref{sec:numerical}).

\begin{figure*}[tp]\centering
\includegraphics[width=0.88\linewidth]{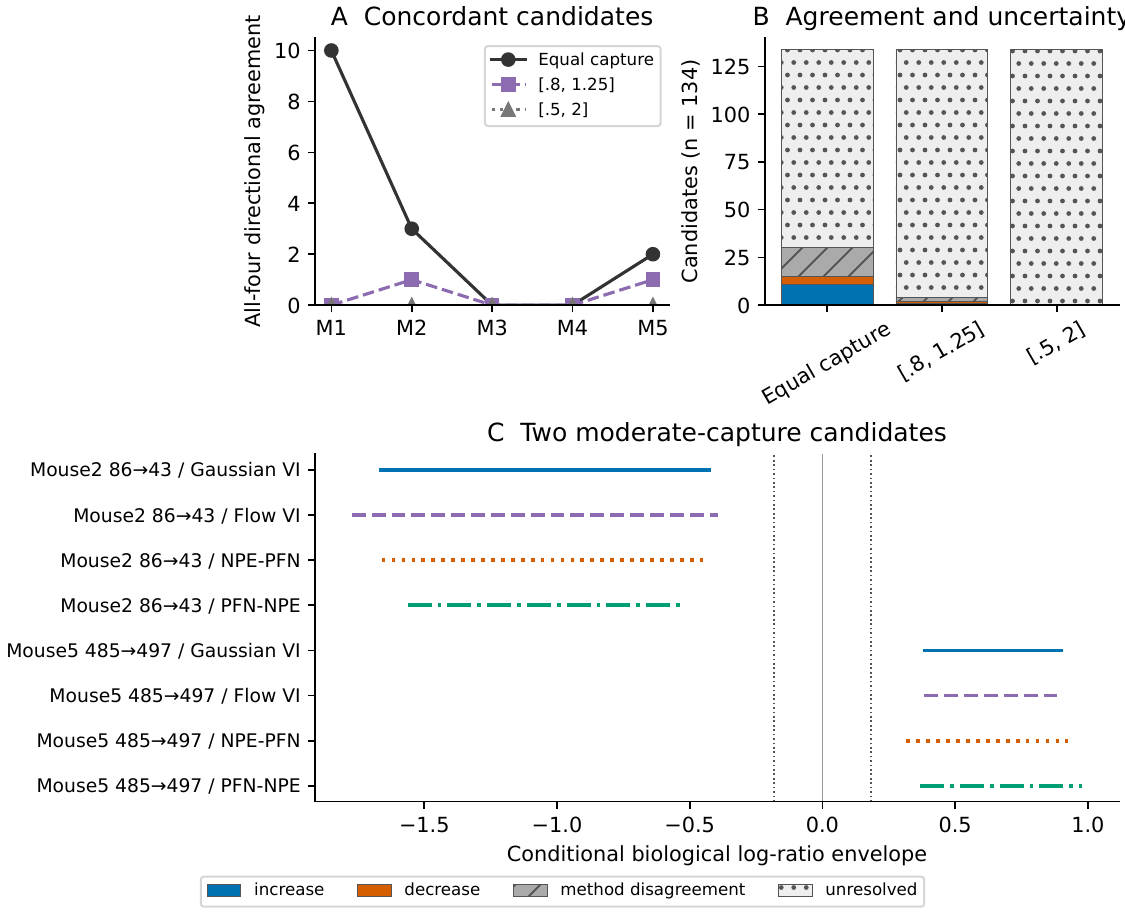}
\caption{Temporal NB contrasts for the 134 tracking-concordant candidates. (A) Counts of candidates with directional agreement across all four methods, by mouse and assumed capture range. (B) A pair is unresolved when no method's interval supports a directional change. Different classifications across methods are labeled disagreement. (C) Capture-adjusted bounds for the biological log-rate ratio in the two pairs with agreement under $q\in[0.8,1.25]$. These bounds expand each 90\% recorded-rate interval over the assumed capture range. Dotted lines mark the 1.2-fold change criterion.}\label{fig:five-nb-candidates}
\end{figure*}

We next asked how often the four methods agreed on a rate change as the capture bounds varied. Among the 134 tracking-concordant pairs, all four 90\% intervals supported the same direction beyond the 1.2-fold margin for 15 pairs at $q=1$, two at $q\in[0.8,1.25]$ and zero at $q\in[0.5,2]$ (Fig.~\ref{fig:five-nb-candidates}). These counts measure agreement between the four posterior approximations under the temporal NB model. Nominal 90\% coverage for rate contrasts was not established in the simulations (Section~\ref{sec:common174-change}). Neither agreement between methods nor allowing for missed spikes corrects errors in the posterior approximations.

At $q\in[0.8,1.25]$, the methods indicated a decrease for the proposed match between Mouse2 unit 86 on the first day and unit 43 on the second day. They indicated an increase for Mouse5 units 485 and 497, respectively. The Mouse2 result was sensitive to the model and analysis window. Its second-day unit's quality summary covered only part of the 29-minute window, and concatenated sorting supported the pair under four of nine settings. Under the Gaussian AR model, this pair did not meet the same-direction change criterion across both priors and windows.

\begin{figure}[tp]\centering
\includegraphics[width=0.92\columnwidth]{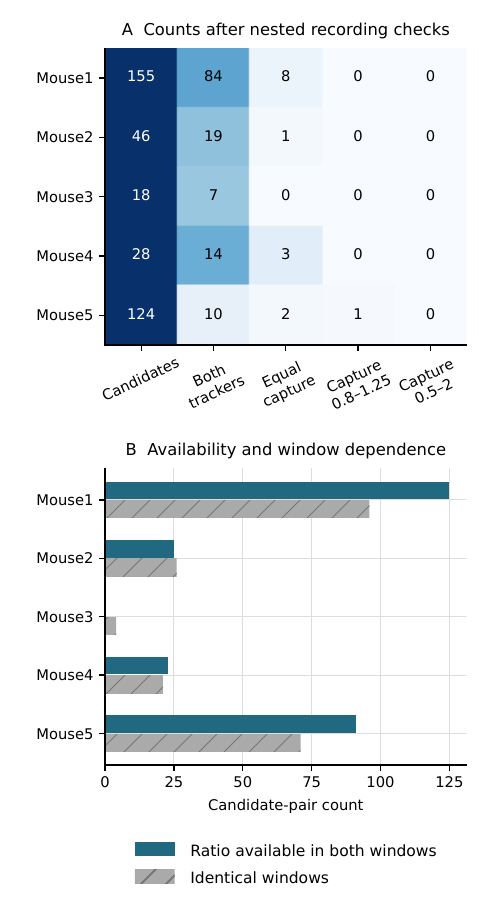}
\caption{Gaussian AR results for 371 candidate pairs. (A) All candidates, those supported by both trackers, and concordant pairs meeting the change criterion under each capture assumption. Intervals must indicate the same direction of change under both priors in the full and quality-overlap windows. (B) Pairs with reportable ratios in both windows and pairs with identical windows. The sole pair meeting the criterion under moderate capture uncertainty had identical windows and concatenated-sorting support under one of nine settings.}\label{fig:five-candidates}
\end{figure}

The temporal NB comparisons use the full 29-minute analysis window. The Gaussian AR analysis also restricts counts to time bins covered by both units' quality summaries and repeats the comparison under two priors. Of the 134 tracking-concordant pairs, the Gaussian AR criterion identified 14 at $q=1$, one at $q\in[0.8,1.25]$ and zero at $q\in[0.5,2]$ (Fig.~\ref{fig:five-candidates}). For the Mouse5 pair meeting the criterion at $q\in[0.8,1.25]$, the full and quality-overlap windows coincided, and concatenated sorting supported the pair under only one of nine settings. For this pair, the prior changes but the two window analyses use the same counts.

Analyzing both windows shows how much interval availability depends on the counts retained. Under the primary prior, 344 of the 371 candidate pairs had reportable rate-ratio intervals for the full-window comparison. The quality-overlap comparison yielded reportable intervals for 265 pairs. Rate-difference intervals were available for 371 and 348 pairs, respectively. The full and quality-overlap windows were identical for 218 pairs.

\section{Discussion and conclusions}\label{sec:discussion}
We present a procedure for inferring changes in stationary mean firing rate after neuron tracking. It combines tracking support, uncertainty in the recorded-rate contrast, and sensitivity to missed spikes. In the five-mouse dataset, the procedure identifies which candidate pairs meet the rate-change criterion under each recording assumption. Each rate comparison is therefore accompanied by its tracking evidence and capture assumptions.

This approach builds on methods for tracking neurons and quantifying apparent changes in activity~\citep{wood2008,ventura2009,stevenson2011}. UnitMatch tracks neurons from spike waveforms, leaving firing rates available for subsequent analysis~\citep{vanbeest2025}. \citet{lu2026} benchmarked tracking on the same five-mouse release and examined neural properties in separate macaque recordings. We use the mouse recordings to quantify how assumptions about missed spikes affect firing-rate comparisons. Our thinning control shows that accounting for known spike loss substantially reduces the number of apparent changes for a fixed unit. In the cross-day analysis, the assumed capture range determines which recorded differences support an increase or decrease in biological firing rate, conditional on the proposed neuron match.

The application makes that dependence visible. Among tracking-concordant pairs, 14 met the Gaussian AR criterion at $q=1$, one at $q\in[0.8,1.25]$ and zero at $q\in[0.5,2]$. The pair meeting the criterion at $q\in[0.8,1.25]$ had identical analysis windows and concatenated-sorting support under one of nine settings. The temporal NB analysis identified two pairs at $q\in[0.8,1.25]$. For the Mouse2 pair, the Gaussian AR criterion was not met in every combination of prior and analysis window, and the second-day quality summary covered only part of the window. For the Mouse5 pair, the two analysis windows were identical, and concatenated sorting supported the match under one of nine settings. The two candidates therefore require different qualifications about recording quality and the strength of the proposed match.

The count-model results provide guidance on estimating uncertainty. NB regression improved held-out prediction in every mouse, supporting a model that allows overdispersion. The simulations also showed that arithmetic rate-difference intervals could remain reportable under strong dependence when log-ratio intervals failed the reporting criteria. The difference interval can therefore remain reportable when the ratio interval fails the positivity or precision checks. Both summaries still depend on the observation model.

The posterior comparisons add evidence about the choice of inference method. NPE-PFN and PFN-NPE reuse pretrained transformers for posterior estimation~\citep{vetter2025,pickens2026}. In our temporal NB model, flow VI had the smallest $\mathrm{SW}_2$ to all seven NUTS references that passed the sampling checks. The PFNs had lower mean forecast loss in Mouse4, while VI had the lowest mean in the other four mice. Coverage and width also revealed different behavior across parameters. Transformer intervals for $\rho$ were often nearly as wide as the prior intervals. Across the 64 simulated rate contrasts, all four coverage estimates were below the nominal 90\% level. The 95\% Wilson confidence intervals for three methods included the nominal level, reflecting uncertainty in the coverage estimates from 64 pairs. These results support evaluating posterior accuracy, interval coverage, width and prediction for the particular inference target. Agreement between methods in the mouse application remains a model-based criterion, rather than a calibrated biological-discovery rule.

We examined two days per mouse in one dataset, including Mouse2 recordings used during development. Differences in behavior, spikes assigned from other neurons, sorting errors and uncertainty about how long the probe was actively recording may affect the rate comparisons. Gaussian AR intervals also depend on the observation model, which permits negative counts, and were checked in selected simulations. Longer recordings with behavioral covariates and measured spike capture would help assess how widely the findings apply. Longer forecast horizons and total computational cost remain to be assessed. The present analysis provides estimates of stationary mean firing-rate change with explicit recording assumptions.

\section*{Data availability}
The five-mouse recordings are available from the UnitMatch example dataset on Figshare, version 2~\citep{unitmatchdata2024}.

\section*{Code availability}
Analysis code and processed results are hosted on \href{https://github.com/nilson01/neural-firing-rate-uncertainty}{GitHub}.

\section*{Funding}
This research received no funding.

\section*{Declaration of competing interests}
The authors declare no competing interests.

\begin{samepage}
\section*{Ethics statement}
No new animal procedures were performed. \citet{vanbeest2025} report the procedures and approvals for the original recordings.
\par\end{samepage}

\FloatBarrier


\begin{thebibliography}{99}

\bibitem[Archer et al.(2015)]{archer2015}
E. Archer, I. M. Park, L. Buesing, J. Cunningham and L. Paninski.
\emph{Black box variational inference for state space models}.
arXiv preprint arXiv:1511.07367, 2015.
\href{https://arxiv.org/abs/1511.07367}{arXiv:1511.07367}.

\bibitem[Barnett et al.(2016)]{barnett2016}
A. H. Barnett, J. F. Magland and L. F. Greengard.
\emph{Validation of neural spike sorting algorithms without ground-truth information}.
Journal of Neuroscience Methods 264, 65--77, 2016.
\href{https://doi.org/10.1016/j.jneumeth.2016.02.022}{doi:10.1016/j.jneumeth.2016.02.022}.

\bibitem[Betancourt(2016)]{betancourt2016}
M. Betancourt.
\emph{Diagnosing Suboptimal Cotangent Disintegrations in Hamiltonian Monte Carlo}.
arXiv preprint arXiv:1604.00695, version 1, 2016.
\href{https://arxiv.org/abs/1604.00695v1}{arXiv:1604.00695v1}.

\bibitem[Blei et al.(2017)]{blei2017}
D. M. Blei, A. Kucukelbir and J. D. McAuliffe.
\emph{Variational Inference: A Review for Statisticians}.
Journal of the American Statistical Association 112, 859--877, 2017.
\href{https://doi.org/10.1080/01621459.2017.1285773}{doi:10.1080/01621459.2017.1285773}.
\href{https://arxiv.org/abs/1601.00670v9}{arXiv:1601.00670v9}.

\bibitem[Byrd et al.(1995)]{byrd1995}
R. H. Byrd, P. Lu, J. Nocedal and C. Zhu.
\emph{A Limited Memory Algorithm for Bound Constrained Optimization}.
SIAM Journal on Scientific Computing 16(5), 1190--1208, 1995.
\href{https://doi.org/10.1137/0916069}{doi:10.1137/0916069}.

\bibitem[Cario(1996)]{cario1996}
M. C. Cario.
\emph{Modeling and Simulating Time Series Input Processes with ARTAFACTS and ARTAGEN}.
Proceedings of the 1996 Winter Simulation Conference, 207--213, 1996.
\href{https://www.informs-sim.org/wsc96papers/027.pdf}{Winter Simulation Conference 1996}.

\bibitem[Carlson et al.(2014)]{carlson2014}
D. E. Carlson, J. T. Vogelstein, Q. Wu, W. Lian, M. Zhou, C. R. Stoetzner,
D. Kipke, D. Weber, D. B. Dunson and L. Carin.
\emph{Multichannel Electrophysiological Spike Sorting via Joint Dictionary Learning and Mixture Modeling}.
IEEE Transactions on Biomedical Engineering 61(1), 41--54, 2014.
\href{https://doi.org/10.1109/TBME.2013.2275751}{doi:10.1109/TBME.2013.2275751}.

\bibitem[Dickey et al.(2009)]{dickey2009}
A. S. Dickey, A. Suminski, Y. Amit and N. G. Hatsopoulos.
\emph{Single-Unit Stability Using Chronically Implanted Multielectrode Arrays}.
Journal of Neurophysiology 102(2), 1331--1339, 2009.
\href{https://doi.org/10.1152/jn.90920.2008}{doi:10.1152/jn.90920.2008}.

\bibitem[Durkan et al.(2019)]{durkan2019}
C. Durkan, A. Bekasov, I. Murray and G. Papamakarios.
\emph{Neural Spline Flows}.
Advances in Neural Information Processing Systems 32, 2019.
\href{https://arxiv.org/abs/1906.04032}{arXiv:1906.04032}.

\bibitem[Gneiting and Raftery(2007)]{gneiting2007}
T. Gneiting and A. E. Raftery. \emph{Strictly Proper Scoring Rules, Prediction, and Estimation}.
Journal of the American Statistical Association 102(477), 359--378, 2007.
\href{https://doi.org/10.1198/016214506000001437}{doi:10.1198/016214506000001437}.

\bibitem[Goris et al.(2014)]{goris2014}
R. L. T. Goris, J. A. Movshon and E. P. Simoncelli.
\emph{Partitioning neuronal variability}.
Nature Neuroscience 17, 858--865, 2014.
\href{https://doi.org/10.1038/nn.3711}{doi:10.1038/nn.3711}.

\bibitem[Hoffman and Gelman(2014)]{hoffman2014}
M. D. Hoffman and A. Gelman.
\emph{The No-U-Turn Sampler: Adaptively Setting Path Lengths in Hamiltonian Monte Carlo}.
Journal of Machine Learning Research 15, 1593--1623, 2014.
\href{https://jmlr.org/papers/v15/hoffman14a.html}{JMLR 15}.

\bibitem[Hollmann et al.(2025)]{hollmann2025}
N. Hollmann, S. M\"uller, L. Purucker, A. Krishnakumar, M. K\"orfer,
S. B. Hoo, R. T. Schirrmeister and F. Hutter.
\emph{Accurate predictions on small data with a tabular foundation model}.
Nature 637, 319--326, 2025.
\href{https://doi.org/10.1038/s41586-024-08328-6}{\nolinkurl{doi:10.1038/s41586-024-08328-6}}.

\bibitem[Huang et al.(2026)]{huang2026}
Y. Huang et al. \emph{Density-based longitudinal neuron tracking in high-density
 electrophysiological recordings}. Patterns 7, 101590, 2026.
\href{https://doi.org/10.1016/j.patter.2026.101590}{doi:10.1016/j.patter.2026.101590}.

\bibitem[Kingma and Ba(2015)]{kingma2015}
D. P. Kingma and J. Ba.
\emph{Adam: A Method for Stochastic Optimization}.
International Conference on Learning Representations, 2015.
\href{https://arxiv.org/abs/1412.6980}{arXiv:1412.6980}.

\bibitem[Kingma et al.(2016)]{kingma2016}
D. P. Kingma, T. Salimans, R. Jozefowicz, X. Chen, I. Sutskever and M. Welling.
\emph{Improved Variational Inference with Inverse Autoregressive Flow}.
Advances in Neural Information Processing Systems 29, 2016.
\href{https://proceedings.neurips.cc/paper_files/paper/2016/hash/ddeebdeefdb7e7e7a697e1c3e3d8ef54-Abstract.html}{NeurIPS 2016}.

\bibitem[Kipnis et al.(2026)]{kipnis2026}
A. Kipnis, M. Binz and E. Schulz.
\emph{Prior-Amortized In-Context Bayesian Inference for Generalized Linear Mixed-Effects Models}.
arXiv preprint arXiv:2609.24422, version 2, October 19, 2026.
\href{https://arxiv.org/abs/2609.24422v2}{arXiv:2609.24422v2}.

\bibitem[Lee et al.(2026)]{lee2026}
H. D. Lee, A. Jha, S. E. Clarke, M. P. Silvernagel, P. Nuyujukian and S. W. Linderman.
\emph{Stiefel Manifold Dynamical Systems for Tracking Representational Drift}.
bioRxiv preprint, 2026.
\href{https://www.biorxiv.org/content/10.64898/2026.03.07.710319v1}{doi:10.64898/2026.03.07.710319 (v1)}.

\bibitem[Lu et al.(2026)]{lu2026}
H.-Y. Lu, H. M. Stealey, Y. Zhao, C. R. Barnett, E. Contreras-Hernandez and S. R. Santacruz.
\emph{Long-term neuron tracking reveals balance of stability and plasticity in functional properties}.
PLOS One 21(6), e0321830, 2026.
\href{https://doi.org/10.1371/journal.pone.0321830}{doi:10.1371/journal.pone.0321830}.

\bibitem[Mittal et al.(2025)]{mittal2025}
S. Mittal, N. L. Bracher, G. Lajoie, P. Jaini and M. Brubaker.
\emph{Amortized In-Context Bayesian Posterior Estimation}.
arXiv preprint arXiv:2502.06601, version 1, 2025.
\href{https://arxiv.org/abs/2502.06601v1}{arXiv:2502.06601v1}.

\bibitem[Modr\'ak et al.(2025)]{modrak2023}
M. Modr\'ak, A. H. Moon, S. Kim, P. B\"urkner, N. Huurre,
K. Faltejskov\'a, A. Gelman and A. Vehtari.
\emph{Simulation-Based Calibration Checking for Bayesian Computation:
The Choice of Test Quantities Shapes Sensitivity}.
Bayesian Analysis 20(2), 461--488, 2025.
\href{https://doi.org/10.1214/23-BA1404}{doi:10.1214/23-BA1404}.

\bibitem[M\"uller et al.(2022)]{muller2022}
S. M\"uller, N. Hollmann, S. Pineda Arango, J. Grabocka and F. Hutter.
\emph{Transformers Can Do Bayesian Inference}.
International Conference on Learning Representations, 2022.
\href{https://arxiv.org/abs/2112.10510}{arXiv:2112.10510}.

\bibitem[Pandarinath et al.(2018)]{pandarinath2018}
C. Pandarinath, D. J. O'Shea et al.
\emph{Inferring single-trial neural population dynamics using sequential auto-encoders}.
Nature Methods 15, 805--815, 2018.
\href{https://doi.org/10.1038/s41592-018-0109-9}{\nolinkurl{doi:10.1038/s41592-018-0109-9}}.

\bibitem[Pickens et al.(2026)]{pickens2026}
E. Pickens, C. Gohel and S. Satya.
\emph{Pre-trained Tabular Foundation Models as Versatile Summary Networks for Neural Posterior Estimation}.
arXiv preprint arXiv:2605.07765, version 1, 2026.
\href{https://arxiv.org/abs/2605.07765v1}{arXiv:2605.07765v1}.

\bibitem[Pillow et al.(2013)]{pillow2013}
J. W. Pillow, J. Shlens, E. J. Chichilnisky and E. P. Simoncelli.
\emph{A Model-Based Spike Sorting Algorithm for Removing Correlation Artifacts
in Multi-Neuron Recordings}. PLOS ONE 8, e62123, 2013.
\href{https://doi.org/10.1371/journal.pone.0062123}{doi:10.1371/journal.pone.0062123}.

\bibitem[Reuter et al.(2025)]{reuter2025}
A. Reuter, T. G. J. Rudner, V. Fortuin and D. R\"ugamer.
\emph{Can Transformers Learn Full Bayesian Inference in Context?}
Proceedings of the 42nd International Conference on Machine Learning,
Proceedings of Machine Learning Research 267, 51531--51582, 2025.
\href{https://proceedings.mlr.press/v267/reuter25a.html}{PMLR 267}; \href{https://arxiv.org/abs/2501.16825v2}{arXiv:2501.16825v2}.

\bibitem[Rule et al.(2019)]{rule2019}
M. E. Rule, T. O'Leary and C. D. Harvey.
\emph{Causes and consequences of representational drift}.
Current Opinion in Neurobiology 58, 141--147, 2019.
\href{https://doi.org/10.1016/j.conb.2019.08.005}{doi:10.1016/j.conb.2019.08.005}.

\bibitem[SciPy developers(2026)]{scipysobol}
SciPy developers. \emph{scipy.stats.qmc.Sobol reference documentation}.
SciPy reference manual, accessed October 30, 2026.
\href{https://docs.scipy.org/doc/scipy/reference/generated/scipy.stats.qmc.Sobol.html}{Sobol engine documentation}.

\bibitem[Stevenson et al.(2011)]{stevenson2011}
I. H. Stevenson, A. Cherian, B. M. London, N. A. Sachs, E. Lindberg,
J. Reimer, M. W. Slutzky, N. G. Hatsopoulos, L. E. Miller and K. P. Kording.
\emph{Statistical assessment of the stability of neural movement representations}.
Journal of Neurophysiology 106, 764--774, 2011.
\href{https://doi.org/10.1152/jn.00626.2010}{doi:10.1152/jn.00626.2010}.

\bibitem[van Beest et al.(2024)]{unitmatchdata2024}
E. van Beest, C. Bimbard, J. Fabre, F. Takacs, P. Coen, A. Lebedeva,
K. Harris and M. Carandini.
\emph{UnitMatch Demo - data}.
figshare dataset, version 2, 2024.
\href{https://doi.org/10.6084/m9.figshare.24305758.v2}{\nolinkurl{doi:10.6084/m9.figshare.24305758.v2}}.

\bibitem[van Beest et al.(2025)]{vanbeest2025}
E. H. van Beest, C. Bimbard et al.
\emph{Tracking neurons across days with high-density probes}.
Nature Methods 22, 778--787, 2025; published online September 27, 2024.
\href{https://doi.org/10.1038/s41592-024-02440-1}{\nolinkurl{doi:10.1038/s41592-024-02440-1}}.

\bibitem[Vehtari et al.(2021)]{vehtari2021}
A. Vehtari, A. Gelman, D. Simpson, B. Carpenter and P.-C. B\"urkner.
\emph{Rank-normalization, Folding, and Localization: An Improved $\widehat R$ for Assessing Convergence of MCMC (with discussion)}.
Bayesian Analysis 16(2), 667--718, 2021.
\href{https://doi.org/10.1214/20-BA1221}{doi:10.1214/20-BA1221}.

\bibitem[Ventura(2009)]{ventura2009}
V. Ventura. \emph{Traditional waveform based spike sorting yields biased rate code estimates}.
Proceedings of the National Academy of Sciences 106, 6921--6926, 2009.
\href{https://doi.org/10.1073/pnas.0901771106}{doi:10.1073/pnas.0901771106}.

\bibitem[Vetter et al.(2025)]{vetter2025}
J. Vetter, M. Gloeckler, D. Gedon and J. H. Macke.
\emph{Effortless, Simulation-Efficient Bayesian Inference using Tabular Foundation Models}.
Advances in Neural Information Processing Systems 38, 2025.
\href{https://arxiv.org/abs/2504.17660v2}{arXiv:2504.17660v2}.

\bibitem[Wilson(1927)]{wilson1927}
E. B. Wilson. \emph{Probable Inference, the Law of Succession, and Statistical Inference}.
Journal of the American Statistical Association 22(158), 209--212, 1927.
\href{https://doi.org/10.1080/01621459.1927.10502953}{doi:10.1080/01621459.1927.10502953}.

\bibitem[Wood and Black(2008)]{wood2008}
F. Wood and M. J. Black. \emph{A nonparametric Bayesian alternative to spike sorting}.
Journal of Neuroscience Methods 173, 1--12, 2008.
\href{https://doi.org/10.1016/j.jneumeth.2008.04.030}{doi:10.1016/j.jneumeth.2008.04.030}.
\end{thebibliography}
\end{document}